\documentclass[aps,prc,showpacs,amssymb,amsmath,amsfonts,nofootinbib,floatfix]{revtex4-2}
\usepackage{graphicx}
\usepackage{lineno}
\makeatletter
\def\p@subsection{}

\def\p@subsubsection{}
\makeatother
\usepackage{natbib}
\usepackage[usenames]{color}
\usepackage{epsfig,graphicx}
\usepackage{epstopdf}
\usepackage{amssymb,amsmath,eqparbox}
\usepackage{subfig}
\usepackage{epstopdf}
\usepackage{enumitem}
\usepackage{comment}
\usepackage[bookmarks={false}]{hyperref}
\usepackage{caption}
\definecolor{grey}{rgb}{0.9,0.9,0.9}

\definecolor{black}{rgb}{0,0,0}

\hypersetup{pdfstartview={XYZ null null 1.}}

\newcommand{\be}{\begin{eqnarray}}
\newcommand{\ee}{\end{eqnarray}}
\newcommand{\bc}{\begin{center}}
\newcommand{\ec}{\end{center}}

\newcommand{\Real}{\text{Re}}
\newcommand{\Imag}{\text{Im}}

\newcommand{\RudjerBoskovic}{Rudjer Bo\v{s}kovi\'{c} Institute, Bijeni\v{c}ka cesta 54, P.O. Box 180, 10002 Zagreb, Croatia}
\newcommand{\Tesla}{Tesla Biotech d.o.o., Mandlova 7, 10000 Zagreb, Croatia}

\begin{document}

\allowdisplaybreaks

\title{Complete set of observables in pseudo-scalar meson photoproduction \\ Controversy solved}
\author{A.~{\v{S}}varc}\email[Corresponding author: ]{svarc@irb.hr}
\affiliation{\RudjerBoskovic,\Tesla}

\date{\today}

\begin{abstract}
\vspace*{1.cm}
The long-standing debate over whether the complete set of observables in pseudo-scalar meson photoproduction consists of eight or merely four elements continues to persist. From the perspective of amplitude analysis, it is argued that all eight observables are necessary to completely determine the others. On the other hand, proponents of partial-wave analysis, working with theoretically precise data of infinite accuracy, claim that only four observables are needed. However, this claim is not acceptable from an experimental viewpoint, as all data in the real world contain some uncertainty. This paper illustrates that the controversy is artificial and is due to additional mathematical assumptions used in partial-wave analysis. Our research advances this discussion by moving from  exact synthetic numerical data to also synthetic, but more realistic data in partial-wave analysis and shows that the claimed reduction in observables is unjustified. Consequently, the final conclusion is that the complete set of observables in pseudo-scalar meson photoproduction, whether using amplitude analysis or partial-wave analysis with practical data, must consist of eight observables.
\end{abstract}

\maketitle

\section{Introduction}
It is imperative to define what is  meant by the complete set of observables, often referred to as a complete experiment. An early definition of this concept was provided by Barker, Donnachie, and Storrow~\cite{Barker1975}, stating: "It is well known that we need 7 measurements to determine the amplitudes up to an overall phase and up to discrete ambiguities," a definition subsequently adopted by Keaton and Workman~\cite{Workman1996}, Chiang Tabakin~\cite{Chiang1997}, and Nakayama~\cite{Nakayama2019}. It has further been demonstrated in \cite{Chiang1997} that all discrete ambiguities can be resolved if an eighth, judiciously selected, observable is included. These analyses, relying on the amplitude analysis technique, are in complete concordance; the complete set of observables is formed by 8 well-chosen observables. Nonetheless, it is crucial to emphasize that this definition does not yield only one definitive set of 4 complex amplitudes; it yields an infinite multitude of them, attributable to the overall phase ambiguity. Therefore, the answer to this problem is inherently non-unique. A specific set of 4 complex amplitudes derived from a chosen set of observables can be multiplied by an unknown phase $e^{i \phi(W,\theta)}$ dependent on energy and angle , modifying all real and imaginary components, yet preserving the same set of observables. This phenomenon is known as the continuum ambiguity effect~\cite{Atkinson1973,Atkinson1985,Bowcock1975,Svarc2018},  and it is essential for our analysis.
\\ \\ \indent
Complete set of observables issues are  also examined in  partial wave analysis,  a well-established methodology for data examination within the domain of experimental physics~\cite{MartinSpearman}; however, the conclusions differed. A substantial scholarly attention has been devoted to revealing the complete set of observables when partial wave analysis is employed. Significant investigative efforts in this domain have been undertaken by the  Mainz/Bonn/GWU collaborations~\cite{Wunderlich2014,Workman2017,Wunderlich2019}. However, their findings diverge from conventional amplitude analysis outcomes: instead of the requisite 8 observables identified by amplitude analysis, they assert that only 5—even potentially 4—observables are necessary when employing partial wave analysis~\cite{Workman2017,Wunderlich2019}. This discrepancy presents a notable challenge to the experimental community's consensus. It raises a critical inquiry: is the acquisition of data from 8 separate measurements indispensable, or can a precise measurement of merely 4 suffice? This question persists unanswered, and it is the objective of this paper to address this issue.
\newpage
To get a convincing answer it is essential to thoroughly examine the conditions under which both approaches work. In the amplitude analysis, the objective is to determine seven real parameters, consisting of four absolute values and three relative angles of four complex amplitudes, at a fixed energy and angular point \{W, \text{Cos}$\theta$\}, without introducing any correlations between adjacent points. It is important to repeat that this analysis is conducted "up to an overall phase." Consequently, the resulting four complex amplitudes are not uniquely defined, and the solution to this problem is an infinite set of four complex amplitudes. The problem's uniqueness is established only upon the introduction of continuum ambiguity considerations and fixing the overall phase by incorporating all other open channels.
Conversely, the technique of partial wave analysis presupposes from the outset that partial wave analysis is feasible, necessitating the existence of a unique set of four complex amplitudes. This issue has been extensively discussed in ref.~\cite{Svarc2018}, and in Chapter~1.5 of ref.~\cite{Wunderlich2019}, where it was explicitly stated: "The problem of the unknown overall phase therefore blocks the way from the CEA to multipoles." As a potential resolution to this issue, it has been suggested to employ a truncated partial wave analysis, where angular momentum is confined to a certain finite cut-off angular momentum value, essentially fixing the angular part of the overall phase \footnote{Truncating partial wave analysis means that  out of infinite number of possible solutions for various overall phases  we fix the overall phase to the value for which all multipoles higher than $L_{max}$ vanish.}.  Technically, the continuity was implemented by fixing energy dependent part of the overall phase by fixing one of the multipole phases.  They have chosen Re$(E_{0+}) > 0 $ and Im($E_{0+})=0$. Their assertion was, following the works of Grushin~\cite{Grushin1989}, that fixing one multipole phase suffices to fix the overall phase in truncated partial wave analysis. Thus, fixing the overall phase through truncation constitutes the primary significant difference. The second difference is introducing analyticity in angle. The angular behavior of the four complex amplitudes is described by partial wave expansion of CGLN amplitudes~\cite{CGLN1957} given in Appendix, so the reaction amplitudes exhibit analyticity concerning the angular variable. In other words, point-to-point continuity is imposed. In conclusion, the fundamental assumptions of amplitude versus partial wave analysis differ, leading to the expectation that additional assumptions made for partial wave analysis will influence the analysis. Furthermore, the analytical techniques employed differ; amplitude analysis bases its conclusions on general considerations, searching for only seven real numbers, with data and measurements largely unneeded. In contrast, partial wave analysis relies on data fitting rather than general considerations; thus, the selection of the data base and numerical aspects of the problem will significantly impact the outcome. It is imperative to ensure that our final conclusions remain unaffected by the particularities of partial wave analysis, and are fully realistic irrespective of the assumptions made in either fitting or database selection.
\section{Complete set of observables in truncated partial wave analysis with fixed multipole phase}

 Let me delineate the methodology employed in refs.~\cite{Wunderlich2014,Workman2017,Wunderlich2019}. The procedure adopted in these studies is as follows:
 \begin{enumerate} [leftmargin=2.cm,noitemsep]
     \item a fixed set of multipoles is selected at a specified energy with fixed cut-off angular momentum;
     \item  precise numerical values of all observables for the selected energy and prescribed cut-off angular momentum are generated; 
     \item the selected set of numerical observables is fitted with a suitable formula at a specified energy and chosen cut-off angular momentum to identify the minimal set of observables required for the unique restoration of the initial value.
 \end{enumerate}  
Thus, their conclusions are not derived through analytical means but are inferred from the numerical analysis of infinite precision (unrealistically ideal) synthetic data for finite angular momentum. As the number of parameters  increases with the order of truncation, the fitting process becomes increasingly slow; hence, they limit their conclusions to $L_{max} = 2$. Therefore, two additional assumptions are made: partial wave analysis is confined to a truncated partial wave analysis with relatively low cut-off angular momentum, and the fitted observables are numerical observables of infinite precision.
 \\ \\ \indent
 This method prompts two further questions: Firstly, in what manner is stability ensured in truncated partial wave analysis, given the established fact that unconstrained partial wave analysis, due to continuum ambiguity, fails to produce continuous outcomes? Secondly, how are the numerical challenges effectively managed, especially since the objective involves locating the exact zero of the minimized function—a task recognized for its significant numerical sensitivity?
\\ \\ \indent
In the truncated partial wave analysis as discussed in refs.~\cite{Wunderlich2014,Workman2017,Wunderlich2019}, the continuity of solution has been ensured by adhering to the approach proposed by Grushin in 1989~\cite{Grushin1989}. This method essentially involves designating the phase of a selected multipole to be real. This process is thoroughly detailed in Chapter 5: "Numerical truncated partial wave analysis" of ref.~\cite{Wunderlich2019}, wherein it is explicitly mentioned that they employ "phase constrained" multipoles $\mathcal{M}^{C}_l$, specified by setting the phase of $E_{0+}$ to zero: $\rm{Re}[E_{0+}] > 0$, and $\rm{Im}[E_{0+}] = 0$. Consequently, they utilized $E_{0+}$ in place of $M_{1-}$ multipole, as recommended in Grushin's work~\cite{Grushin1989}. This methodology successfully achieved solution continuity and corroborated Omelaenko's finding that a carefully selected set of four observables constitutes a complete set~\cite{Omelaenko1981}. This outcome posed a conundrum for experimentalists, as all the underlying assumptions leading to this conclusion appeared quite plausible. Does this imply that only four observables are necessary, rather than eight, to comprehensively describe the process?
\\ \\ \indent
The numerical partial wave analysis of data with infinite precision, as performed by the Mainz/Bonn/GWU groups for $L_{max}=2$, has been successfully replicated, ensuring continuity through the enforcement of a single constraint: $E_{0+}$ the multipole must be real and positive. Consequently, it is necessary to derive 8 multipoles by fitting 15 real parameters to the numerically precise data generated by the exact formula for $L_{max}=2$. This endeavor posed a complex mathematical challenge, given the goal of achieving 15 parameters while fitting a minimization function that equals precisely zero. Despite the numerical intricacies, I have successfully reproduced the outcomes detailed in refs.~\cite{Wunderlich2014,Workman2017,Wunderlich2019}, thereby corroborating all the results presented in Table III of ref.~\cite{Workman2017} using the Mathematica code, akin to the implementation in the original investigations. Under these conditions, a set of four carefully selected observables constitutes a complete set.
\\ \\ \indent
As new codes tend to be faster than the original codes from the Mainz-Bonn-GWU group, I have confirmed these results for  $L_{max} = 3$, with the ability to go even higher in $L$ if needed. 
\\ \\ \indent
However, let me stress that their conclusions were made for truncated partial wave analysis, and this means that the angular part of the overall phase was fixed by choosing the truncation order.  In the real world, truncation in partial wave analysis is only an approximation, and we have an infinite number of terms instead of finite ones, so the angular part of the overall phase is not fixed, and it has to be fixed in addition. So, we have to fix amplitude phase, and not multipoles phase only.  
\section{Complete set of observables in infinite partial wave analysis with fixed overall amplitude phase}
To be as similar as possible to Mainz-Bonn-GWU studies, I have chosen to fix the phase of F1 CGLN amplitude, and decided to put it to zero requiring that Re[$F1(W,\theta)] > 0 $ and Im[$F1(W,\theta)]  =  0$. 
To achieve this, I have devised the methodology for independently varying a set of multipoles as fitting parameters while preserving the constancy of the overall phase. As the overall phase is an unmeasurable quantity, and following ref.~\cite{Svarc2024}, from now on I shall in our analysis identify the overall phase with the CGLN F1 amplitude phase.   This necessitated the imposition of a supplementary set of constraints, which are more stringent than the mere fixation of a single multipole phase. 

An efficient approach to maintaining the overall phase invariance was identified in the CGLN representation of the reaction amplitudes. Specifically, among the various representations of reaction amplitudes (CGLN, helicity, transversity, etc.), this is the sole representation where all amplitudes can be articulated as an expansion only in powers of cosine, without the interference from any sine-dependent factors that are present in other representations. So I  rewrite the formulas~\ref{eq:MultExpF1-F4} in the  following way:
\be  \label{CGLN}
F_i(W, {\rm Cos \, }  \theta) &=& \sum_{l=0}^{L_{max}} a_l^i(W) \, {\rm Cos \, } \theta^l \, \, \, \, \, \,   i=1,...,4
\ee
where  $a_l^i(W)$ represent coefficients that exhibit linear dependency on all employed multipoles.  
\newpage
In the case of $L_{max}= 2$, the three coefficients associated with each of the four amplitudes are specified as follows:
\be \label{coeff}
a_0^1(W) &= & E_{2-} + E_{0+}- \frac{3}{2} \, E_{2+} + 3 M_{2-} -3 M_{2+} \\ \nonumber
a_1^1(W) &= & 3 (E_{1+} +  M_ {1+}) \\  \nonumber
a_2^1(W) &= & 15  (\frac{1}{2} \, E_{2+} +  M_{2+} )  \\  \nonumber \\  \nonumber
a_0^2(W) &= & M_{1-}  + 2 M_{1+}  \\   \nonumber 
a_1^2(W) &= & 3(2  M_{2-} + 3 M_{2+})   \\  \nonumber  \\  \nonumber
a_0^3(W) &= &  3 ( E_{1+} - M_{1+})  \\    \nonumber
a_1^3(W) &= & 15( E_{2+}-M_{2+}  )   \\ \nonumber  \\  \nonumber
a_0^4(W) &= & 3 (- E_{2-} - E_{2+} - M_{2-} + M_{2+})   \\    \nonumber
\ee
 Now it is trivial to implement the condition Im[$F1(W,\theta)]  =  0$.  We demand that each coefficient of expansion in Cos$\theta$ must be equal to zero. So, I get:
 \be
 {\rm Im} \, \, \left[a_0^1(W)\right] & = 0 \\ 
  {\rm Im} \, \, \left[a_1^1(W)\right] & = 0  \nonumber  \\ 
   {\rm Im} \, \, \left[a_2^1(W)\right] & = 0 \nonumber 
 \ee
This yields a system of three constraint equations incorporating 8 imaginary parts that define the eight utilized multipoles. Solving this system permits the removal of three real parameters from the original set of 16, consequently leaving 13 parameters free. Within this framework, the F1 phase persists in its invariance when all 13 remaining multipoles are permitted to vary freely, although the absolute value may fluctuate without constraints.
\\ \\ \indent
Accordingly, the fit was conducted using three constraints rather than a single one, as was the case when a single multipole phase was fixed. This modification resulted in a reduction of the number of observables constituting a complete set, from four to three. 
\\ \\ \indent
Let me summarize.  Both aforementioned analyses revealed that we encounter five ambiguities when restoring multipoles from a complete set of  observables. Three ambiguities are sign ambiguities  originating in the fact that all observables are generated by a set of formulas which include sines and cosines of three relative angles,  and can be eliminated by measuring a sufficient number of observables (three of them). However, two phase ambiguities (ambiguities in energy and angle) are genuine ambiguities and cannot be eliminated in a single channel measurement; they have to be fixed by assumption. In truncated partial wave analysis, the energy dependent part is fixed by hand introducing the assumption that the phase of $E_{0+}$ multipole vanishes (\Real $E_{0+} > 0$ and \Imag $E_{0+} = 0$), and  the angular dependent part is fixed by choosing the truncation order. So, we are left with four ambiguities. Three sin/cos ambiguities are solved by extending the number of measured observables to three, and the angular phase ambiguity is solved by adding one more observable to the measured set~\footnote{Observe that in this formalism angular phase ambiguity is for truncated PWA transferred to observables. This seems unnatural as phase ambiguity should not be resolved by measuring observables, but in this procedure this happens.  Remember that we are fitting synthetic observables which have to be generated somehow. And in generating them,  when we fix the energy dependent phase ambiguity by a free choice, different levels of truncation  produce different sets of observables as truncation picks only one solution out of infinite number of them in which all  multipoles higher then cut-off  angular momentum $L_{max}$.}. For infinite PWA, the angular dependence of phase ambiguity is not fixed, so we have to do it by hand too. So, we choose to define that the phase of $F1(W, \theta)$ CGLN amplitude always vanishes ($\Real F1(W, \theta) > 0 $ and $\Imag F1(W, \theta) = 0 $ for each possible angle $\theta$). So, we have to choose a set of only three observables to eliminate sin/cos ambiguity. 
\\ \\ \indent
It is essential to recognize that full analyticity has not yet been implemented due to the absence of analyticity in energy. Should energy analyticity be established, each observable would possess a unique value with respect to  sign ambiguity. This is because dispersion relations would resolve the indefinite sign associated with the cosine and sine elements within the formulas of the observables.
\\ \\ \indent
Consequently, I contend that incorporating energy analyticity will resolve all sin/cos  ambiguities regarding the definitions of observables, leaving a single observable necessary to constitute a complete set. Naturally, it is essential to measure this at an adequate number of energy points, aligning with the principles of analytic function theory.

\section{Complete set of observables with analyticity in energy explicitly introduced}

In order to examine this matter, I rely on my computational codes~\footnote{As I said, we are limited to numeric analysis only.} for the analysis  which are outlined before for $L_{max} =2$, and in addition incorporate analyticity in energy. The model for implementing the analyticity is based on the application of the Pietarinen expansion technique~\cite{Pietarinen}. I employ the definition of CGLN amplitudes given by Eq.(\ref{CGLN}) as it utilizes the expansion in Cos $\theta$ only,  and assume that all coefficients can be expanded in energy using the Pietarinen series.
\be \label{Piet}
a_l^i(W) &=& \sum_{n=1}^{N^i} c_{ln}^i  \, Z(W,\alpha_l^i,\beta_l^i)^n,  \qquad \qquad \qquad  i=1,...,4   \\
Z(W,\alpha_l^i,\beta_l^i) &=&  \frac{\alpha_l^i-\sqrt{\beta_l^i-W}}{\alpha_l^i+\sqrt{\beta_l^i-W}}  \nonumber
\ee
where $N^i$ denotes the order of the Pietarinen expansion, and $c_{ln}^i $, $\alpha_l^i$, and $\beta_l^i$ represent the real Pietarinen coefficients, which vary for each amplitude and angular momentum index. Consequently, analyticity in energy is facilitated through the Pietarinen expansion.
It is now evident that for each specified energy, the relative sign of real and imaginary parts is definitively determined by the Pietarinen expansion, thereby resolving all possible ambiguities.
 The fitting parameters have transitioned from multipoles to Pietarinen coefficients $c_{ln}^i $, $\alpha_l^i$, and $\beta_l^i$. 
 \\ \\ \indent
 To enable numeric analysis, I have simplified the model. I have fixed the Pietarinen parameters $\alpha_l^i$ and $\beta_l^i$, and proceeded to only fit $c_{ln}^i $. Additionally, rather than summing over the entire Pietarinen expansion as indicated in Eq.~(\ref{Piet}), I have limited our consideration of analyticity in energy to a single term.
\be
n=1 \quad {\rm for} \quad i=1; \qquad n=2 \quad {\rm for} \quad  i=2 &  ; & \qquad  n=3 \quad {\rm for} \quad i=3 ;\qquad  n=4 \quad {\rm for} \quad  i=4
\ee
Under this assumption, the preservation of the overall phase occurs automatically. It is essential to emphasize the necessity for additional constraints for fixing the overall phase (akin to those previously employed) if the number of Pietarinen terms for a specified angular momentum coefficient $a_l^i(W)$ exceeds one. The current number of fitting parameters $c_{ln}^i$  amounts to eight, comprising three parameters for F1, two for F2, two for F3, and a single one for F4, as referenced in Eq.~(\ref{coeff}). This is anticipated, given that eleven free parameters were previously required to achieve a unique result with sign ambiguity across three relative angles, necessitating three additional constraints to resolve this issue. In circumstances where analyticity in energy was not manifestly imposed, three ambiguities in relative angles existed, which required three observables for resolution. However, with the explicit imposition of analyticity in energy, these three ambiguities have been eliminated.
\\ \\ \indent
Furthermore, for the sake of simplicity, I posit that the strength and threshold parameters in the Pietarinen expansion are identical for all angular coefficients.
\be
     \alpha_l^i = \alpha  \quad  & {\rm and} & \quad \beta_l^i = W_{thr}^{K \Lambda}
\ee
For the benefit of the reader, I present the final set of CGLN amplitudes within this model:
\be
F_1(W, {\rm Cos \, }  \theta) &=&  \left( \frac{\alpha-\sqrt{W_{thr}^{K \Lambda}}-W}{\alpha+\sqrt{W_{thr}^{K \Lambda}}-W} \right)  \left(c_{01}^1 +  c_{11}^1 \, \, {\rm Cos} \theta +  c_{21}^1 \, \, {\rm Cos} \theta ^2  \right) \\
F_2(W, {\rm Cos \, }  \theta) &=&  \left( \frac{\alpha-\sqrt{W_{thr}^{K \Lambda}}-W}{\alpha+\sqrt{W_{thr}^{K \Lambda}}-W} \right)^2  \left(c_{02}^2 +  c_{12}^2 \, \, {\rm Cos} \theta \right) \nonumber \\
F_3(W, {\rm Cos \, }  \theta) &=&  \left( \frac{\alpha-\sqrt{W_{thr}^{K \Lambda}}-W}{\alpha+\sqrt{W_{thr}^{K \Lambda}}-W} \right)^3  \left(c_{03}^3 +  c_{13}^3 \, \, {\rm Cos} \theta \right) \nonumber \\
F_4(W, {\rm Cos \, }  \theta) &=&  \left( \frac{\alpha-\sqrt{W_{thr}^{K \Lambda}}-W}{\alpha+\sqrt{W_{thr}^{K \Lambda}}-W} \right)^4  c_{04}^4  \nonumber 
\ee
The previously employed methodology is repeated: numeric observables of infinite precision are generated and subsequently fitted with the functions that originally produced them. At this stage, the fitting process presents increased complexity numerically, necessitating a greater number of data points and enhanced precision to achieve the final result. However, it is concluded that only a single observable is required to accurately reproduce the specified input.
\\ \\ \indent
Thus, this numerical analysis within a simplified model indicates that a comprehensive set of observables for fully analytical CGLN amplitudes with respect to both angle and energy comprises a single observable. Measurement of the differential cross-section at a sufficient number of points, with no experimental error, will suffice to uniquely determine all four CGLN amplitudes.. 
\\  \\ \indent
And this is indeed a contradiction with amplitude analysis results, which say that you need 8 observables for the same task. And this cannot be true. 
\section{Randomizing numeric set of observables}
\begin{figure}[h!]
\bc
\includegraphics[width=0.3\textwidth]{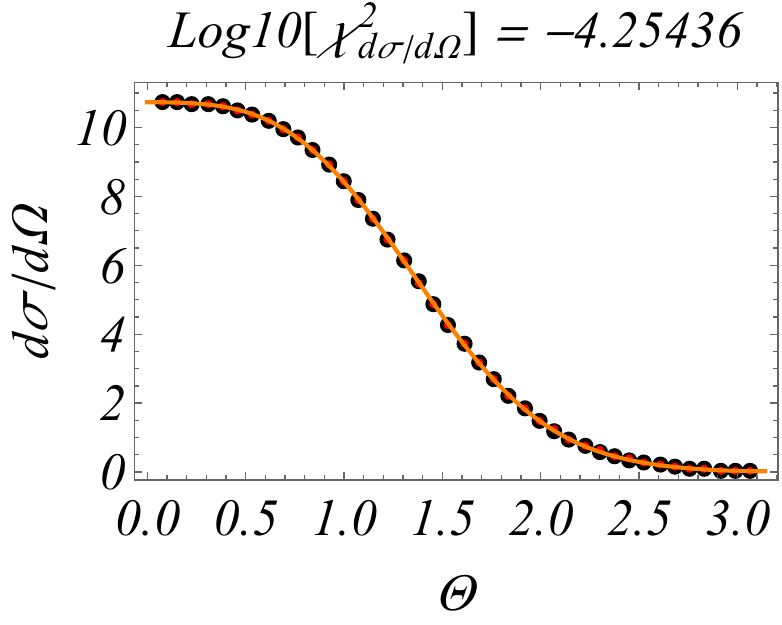} \includegraphics[width=0.3\textwidth]{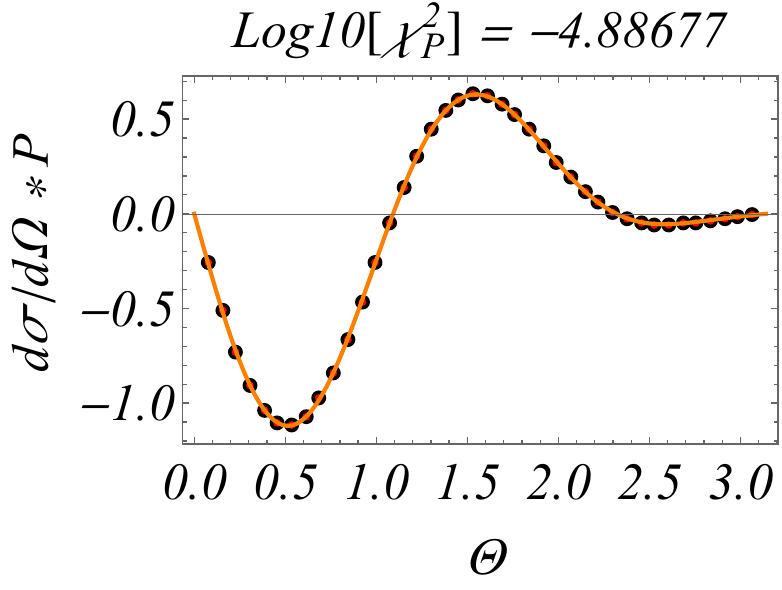}  \includegraphics[width=0.3\textwidth]{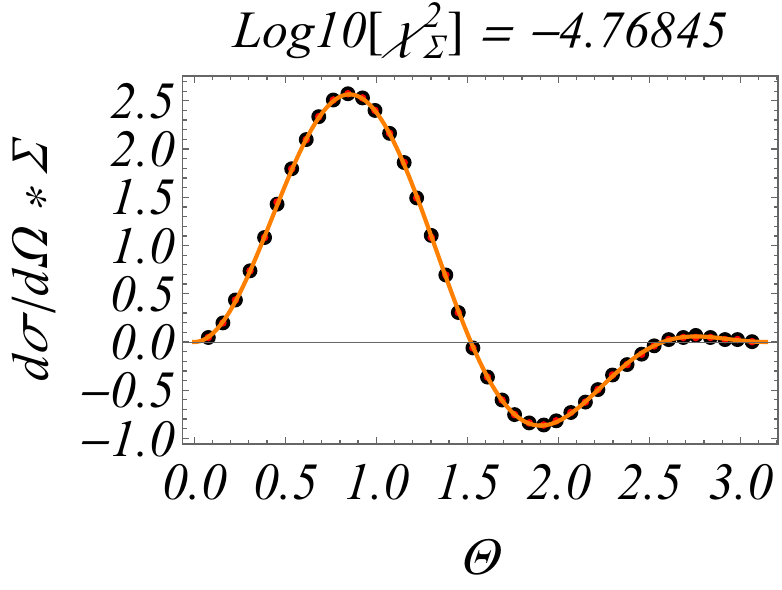} \\ 
\includegraphics[width=0.3\textwidth]{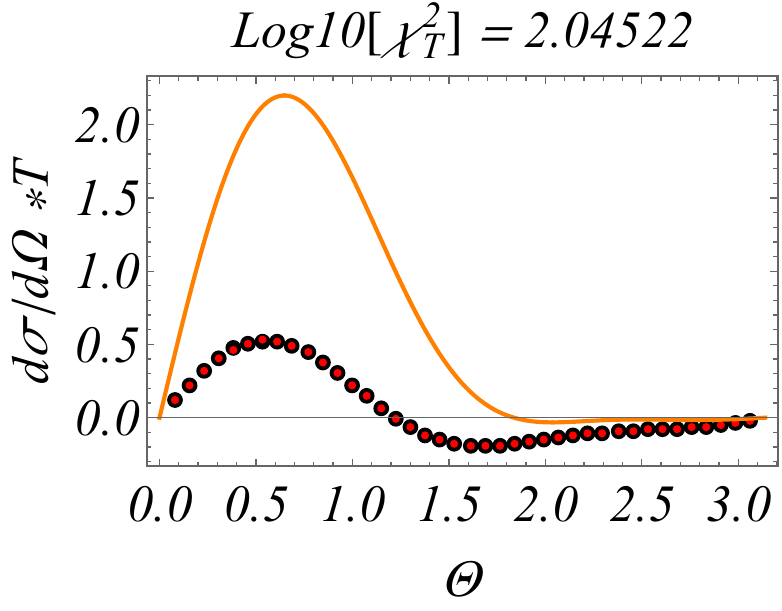} \includegraphics[width=0.3\textwidth]{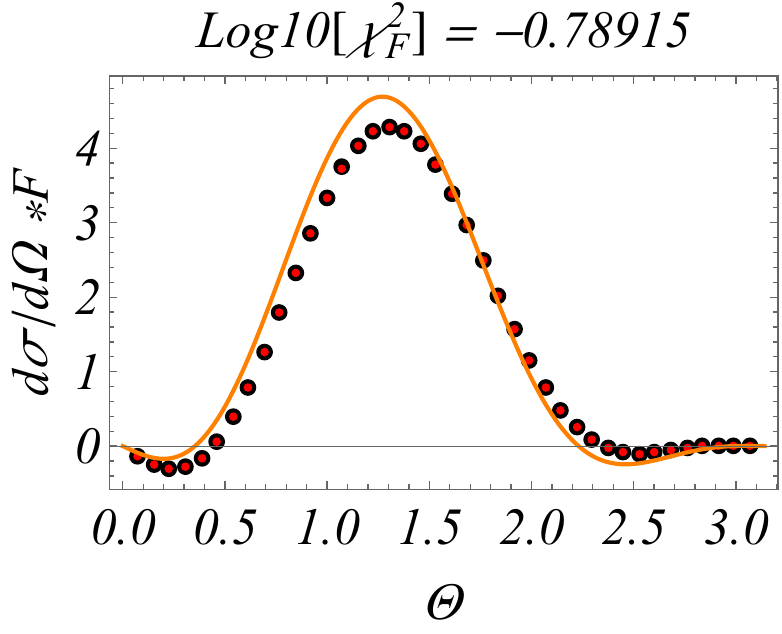}  \includegraphics[width=0.3\textwidth]{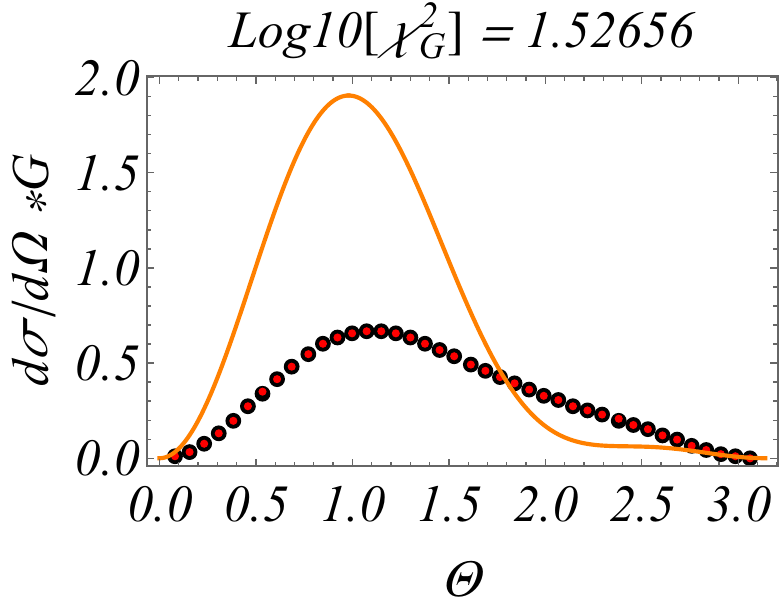}
\caption{\label{Randomized-data}(Color online) Comparison of synthetic numeric data with infinite precision (black circles) at one randomly chosen energy, randomized synthetic data with randomization factor of 0.1 \% (red circles) with the result of the fit to three observables (orange line). First row are fitted observables, and the second row are the predicted values for three randomly chosen observables out of remaining 13. } 
\ec
\end{figure}

It is imperative to emphasize that all prior conclusions, regardless of whether they are made by  fixing $E_{0+}$ multipole phase,  fixing F1(W, Cos$\theta$) amplitude phase, or by introducing manifest analyticity in angle and energy,  have been drawn under the assumption of measurements with negligible experimental error, a condition that is not representative of real-world scenarios. It is significant to note that the authors of the truncated partial wave analyses referenced in~\cite{Wunderlich2014,Workman2017,Wunderlich2019} were cognizant of this limitation and have repeatedly cautioned readers that their research, which relies on data of infinite precision, might be only a consistency exercise. Despite acknowledging this fact, they have not made any further progress on this issue, merely stating that it warrants verification. Consequently, I felt motivated to undertake the subsequent step that  deemed necessary.
\\ \\ \indent
In recognition of the unavoidable presence of experimental error, this study seeks to examine the effects of deviating from the idealistic assumption of data free from error. The standard partial-wave model, characterized by the absence of energy analyticity, was employed to fit the three observables $d\sigma/d\Omega$, $P$, and $\Sigma$. A  randomization was applied to the central values of the synthetic numerical database while maintaining a zero-error baseline. A systematic error of 0.1\% of the maximum value for each observable was incorporated, as well as a relative error of a similar magnitude of 0.1\% of the precise numerical value. This error magnitude, while seemingly trivial and unattainable in practical measurements, is demonstrated in Fig.~(\ref{Randomized-data}). The figure illustrates the non-randomized synthetic numerical data at one randomly chosen energy as black circles, the randomized synthetic data as red circles, and the fitting results as an orange line, with squared-numerical-deviation  values indicated as quantities $\chi^2_{d\sigma/d\Omega,P,\Sigma,T,F,G}$. Although the effects of randomization are scarcely detectable in the figure, it becomes apparent that even with such a slight randomization constant, the predictive accuracy for other, non-fitted observables is significantly reduced when fitting the three observables with full precision.
 Crucially, even with significantly smaller randomization values (several orders of magnitude lower), the fitted multipoles exhibit notable deviations from the initial ones, signifying a loss of completeness. Nonetheless, this impact remains imperceptible within the figures of observables themselves, but remains clearly present when relevant $\chi^2$ values are compared.
\newpage
\section{Conclusions}

I have replicated the truncated partial-wave analysis fits wherein the phase of a single multipole was constrained, as outlined in references \cite{Wunderlich2014,Workman2017,Wunderlich2019}, effectively reproducing the results previously documented even extending the highest truncation order from $L_{max} = 2$ to $L_{max} = 3$ with the possibility to go higher. I have confirmed that in this case the complete set of observables consists of four well-chosen observables.  When the assumption of truncation is banned, and the infinite partial wave analysis is undertaken, the angular part of the overall phase is  not anymore fixed, so we have fixed it in addition. As a result, the number of observables necessary to constitute a complete set was diminished from four to three. Furthermore, by integrating analyticity with respect to energy through a theoretical model, I discerned that the requisite number of observables for a complete set is further reduced from three to one. Thus, I deduce that the reduction in the number of observables from eight in amplitude analysis to one in truncated partial-wave analysis, when analyticity in both angles and energy is explicitly imposed, is attributable to the intrinsic properties of analytic functions when utilizing data sets of infinite precision. I examined the significance of employing data of infinite precision by introducing a version of the fitted data set with minimal randomization and observed the effect's disappearance. This phenomenon is illustrated in the figures, where observables were randomized with a very small randomization coefficient of 0.1 \%. These findings suggest that the reduction in the number of observables from eight in amplitude analysis to one in partial-wave analysis with complete analyticity in both angle and energy, results solely from the employment of idealized numerical observables.
\\ \\ \indent
In conclusion, it has been ascertained that the complete set of observables necessary for truncated partial-wave analysis, when employing realistic data, remains unchanged in comparison to amplitude analysis. The number of observables remains eight.
 \newpage
\begin{acknowledgments}
I am extremely grateful to Ron Workman for numerous discussions and criticisms of various parts of this research. This helped me a lot to prepare this version of the manuscript, for which I hope to be as clear as possible about the main motivation and final results of the research. 
I also express my gratitude to Lothar Tiator and Yannick Wunderlich for their extensive feedback about implicit details of their research that are sometimes not immediately visible to the reader. This also turned out to be instrumental in shaping the current version of this text. 
\end{acknowledgments}  

\appendix
\section{} \label{Appendix}
Partial wave decompositions are introduced through CGLN amplitudes:
\begin{align}
F_{1} \left( W, \theta \right) &= \sum \limits_{\ell = 0}^{\infty} \Big\{ \left[ \ell M_{\ell+} \left( W \right) + E_{\ell+} \left( W \right) \right] P_{\ell+1}^{'} \left( \cos \theta \right) \nonumber \\
 & \quad \quad \quad + \left[ \left( \ell+1 \right) M_{\ell-} \left( W \right) + E_{\ell-} \left( W \right) \right] P_{\ell-1}^{'} \left( \cos \theta \right) \Big\} \mathrm{,} \nonumber  \\
F_{2} \left( W, \theta \right) &= \sum \limits_{\ell = 1}^{\infty} \left[ \left( \ell+1 \right) M_{\ell+} \left( W \right) + \ell M_{\ell-} \left( W \right) \right] P_{\ell}^{'} \left( \cos \theta \right) \mathrm{,}   \label{eq:MultExpF1-F4}  \\
F_{3} \left( W, \theta \right) &= \sum \limits_{\ell = 1}^{\infty} \Big\{ \left[ E_{\ell+} \left( W \right) - M_{\ell+} \left( W \right) \right] P_{\ell+1}^{''} \left( \cos \theta \right) \nonumber \\
 & \quad \quad \quad + \left[ E_{\ell-} \left( W \right) + M_{\ell-} \left( W \right) \right] P_{\ell-1}^{''} \left( \cos \theta \right) \big\} \mathrm{,} \nonumber  \\
F_{4} \left( W, \theta \right) &= \sum \limits_{\ell = 2}^{\infty} \left[ M_{\ell+} \left( W \right) - E_{\ell+} \left( W \right) - M_{\ell-} \left( W \right) - E_{\ell-} \left( W \right) \right] P_{\ell}^{''} \left( \cos \theta \right) \mathrm{.} \nonumber
\end{align}

\end{document}